\documentclass[aps,prx,reprint,onecolumn,superscriptaddress,longbibliography]{revtex4-2}
 
\usepackage{graphicx}
\usepackage{subfigure}
\usepackage{amsmath}
\usepackage{amsfonts}
\usepackage{amssymb}
\usepackage{times,txfonts}
\usepackage{braket}
\usepackage[normalem]{ulem}
\usepackage[colorlinks=true,linkcolor=blue,urlcolor=blue,citecolor=blue]{hyperref}

\begin{document}

\title{The response of a quantum system to a collision: an autonomous derivation of Kubo's formula}


\author{Samuel L. Jacob}
\email{samjac91@gmail.com}
\affiliation{School of Physics, Trinity College Dublin, Dublin 2, Ireland}

\author{John Goold}
\email{gooldj@tcd.ie}
\affiliation{School of Physics, Trinity College Dublin, Dublin 2, Ireland}
\affiliation{Trinity Quantum Alliance, Unit 16, Trinity Technology and Enterprise Centre, Pearse Street, Dublin 2, D02 YN67, Ireland}
\affiliation{Algorithmiq Limited, Kanavakatu 3C 00160 Helsinki, Finland}

\begin{abstract}

We study the response of a quantum system induced by a collision with a quantum particle, using the time-independent framework of scattering theory. After deriving the dynamical map for the quantum system, we show that {it encodes} a non-perturbative response function obeying a general fluctuation-dissipation relation. We show that Kubo's formula emerges autonomously in the Born approximation, where the time-dependent perturbation is determined by particle's evolution through the potential region. 



\end{abstract}

\maketitle{}

\section{Introduction}

Linear response theory describes the response of a physical system to a weak, externally-imposed perturbation \cite{Kubo1957-I}. In particular, it implies that the response of a system in thermal equilibrium is determined by thermal fluctuations -- the so-called fluctuation-dissipation relations \cite{Callen1951,Kubo1966}. The universality and experimental relevance of linear response have made it paramount in the study of classical and quantum systems out of equilibrium, enabling the computation of transport coefficients and susceptibilities, as well as thermodynamic and dynamical properties \cite{Kubo1991,Bruus2004,Calzetta2008,Altland2009,Ciliberto2013,Coleman2015,Banyai2020}. Remarkably, the unveiled connection between linear response and quantum information theory also enables the quantification of multipartite entanglement in equilibrium \cite{Hauke2016,Brenes2020,Scheie2021} and non-equilibrium quantum systems \cite{Pappalardi2017}. Although linear response theory was originally developed for closed systems -- undergoing Hamiltonian or unitary evolution -- several works have been able to extend it to open classical \cite{Agarwal1972,Seifert2010} and quantum \cite{Davies1978,Zanardi2016,Ban2017,Mehboudi2018,Konopik2019,Levy2021,Blair2024} systems. Regarding the latter, one makes use of the theory of open quantum systems to describe the interaction of a quantum system with another system (often playing the role of a thermal environment). The evolution of the open quantum system is then given by a completely positive and trace preserving (CPTP) map \cite{Breuer2007,Alicki2007,Rivas2012}. 

For both open and closed quantum systems, linear response theory considers the perturbation as a time-dependent function or operator. It is understood that such a time-dependence represents a classical system which can be used to drive the system of interest in a controlled fashion \cite{Balian1991}. From a thermodynamic point of view, this classical system constitutes an ideal work source and thermodynamic laws for quantum systems can then be formulated \cite{Alicki1979,Kosloff2013,Goold2016}. In this field, linear response theory already helped elucidate quantum signatures in the power output of quantum heat engines \cite{Brandner2017} and in fluctuations of work \cite{Guarnieri2024}. However, a microscopic approach to the response would describe the driving source as a quantum system in its own right within a time-independent framework. For example, localized massive particles traveling through a time-independent potential can induce an effective time-dependent interaction on the internal degrees of freedom of the particle or scatterer, playing the role of a work source \cite{Jacob2022,Piccione2024}; or alternatively, they can play the role of a heat source if they are sufficiently delocalized \cite{Jacob2021,Tabanera2022,Tabanera2023}.

In this paper, we study the response of a quantum system to a collision with a quantum particle using the time-independent framework of scattering theory. We derive the CPTP map ruling the dynamics of the quantum system and show that the Lamb-Shift Hamiltonian -- {ruling the scattering dynamics at first-order in the scattering amplitudes} -- defines a non-perturbative response function. A fluctuation-dissipation relation is proven which relates the Fourier transform of the response function to a two-point correlation function. We show how Kubo's formula emerges in the Born (first-order) approximation for a localized particle, where the time-dependent perturbation is determined by the particle's evolution through the scattering potential.

Quantum scattering is an essential tool across many realms of physics \cite{Belkic2004,Taylor2006}. Recently, it has been used access thermodynamic fluctuations \cite{Jacob2023,Jacob2024} and finds many applications in quantum transport \cite{Buttiker1992,Nazarov2009,Moskalets2012}, ultracold gases \cite{Inguscio1999,Hohmann2016,Schmidt2018,Schmidt2019,Widera2020,Bouton2021,Adam2022,Nettersheim2023} and condensed matter \cite{Furrer2009,Bramwell2014,Irfan2021,Scheie2021,Laurell2024}. In particular, inelastic neutron scattering has been very successful in condensed matter and can be used to probe density correlations of the scatterer, which are then interpreted \textit{a posteriori} within the linear response formalism \cite{Scherm1972,Berk1993,Furrer2009}. Instead, our approach shows how the latter emerges from the former.  Moreover i) we focus on the dynamics of the quantum system rather than the particle or neutron probe; ii) we do not assume that the latter is delocalized (plane wave approximation); iii) our approach is non-perturbative and the Born approximation is only used to recover Kubo's formula and not as starting point.

The paper is organized as follows. In Section \ref{sec:scattmap} we describe the scattering setup and derive the dynamical map for the quantum system. We also discuss the properties of the map in the light of existing literature. In Section \ref{sec:response}, we show how the map defines a non-perturbative response function which obeys a fluctuation-dissipation relation and show how Kubo's formula is recovered in the Born approximation. After discussing our results, we conclude in Section~\ref{sec:conclusions}. Appendix \ref{app:linearresponse} contains an overview of linear response theory for closed quantum systems. { Appendices~\ref{app:scattmatrixproperties}, \ref{app:qme} and \ref{app:fouriertransform} and pertain only to scattering theory: in the first, we derive important properties of the scattering matrix and amplitudes; in the second, we derive a quantum master equation from the dynamical map, which serves to connect our results with the broader and more familiar literature on quantum master equations; in the third, we discuss the Fourier transforms in the non-perturbative regime of scattering theory.}

\section{The scattering map \label{sec:scattmap}}

\subsection{Setup}

We consider a quantum scattering process between a system $S$ and a particle $P$ with full Hilbert space $\mathcal{H} = \mathcal{H}_S \otimes \mathcal{H}_P$. In a reference frame
co-moving with the center of mass, only the reduced
mass plays a role, but we simplify the treatment
by fixing the position of system $S$ and consider the particle $P$ to be travelling in one direction with associated momentum $\hat{p}$ and position $\hat{x}$ operators. The full Hamiltonian is 
\begin{align}
    \label{hamiltonian}
    \hat{H} = \hat{H}_0 + \hat{V}(\hat{x}) = \hat{H}_S \otimes \hat{\mathbb{I}}_P + \hat{\mathbb{I}}_S \otimes \frac{\hat{p}^2}{2m} + \hat{V}(\hat{x}) \; .
\end{align}
The free Hamiltonian $\hat{H}_0$ is composed of two terms. The first defines the energy of the system $\hat{H}_S \ket{j}=e_j\ket{j}$, where $\{ \ket{j} \}_{j=1}^{N}$ is an orthonormal eigenbasis associated to its discrete energy spectrum $\{ e_j \}_{j=1}^{N}$ and $N = \dim(\mathcal{H}_S)$. The second has a continuous (infinite-dimensional) spectrum and defines the kinetic energy of the particle $\hat{p}^2/2m \ket{p} = E_p \ket{p}$, where $\{ \ket{p} \}$ are improper (non-normalizable) eigenvectors whose position representation are plane waves $\braket{x|p} = \exp( i p x / \hbar) / \sqrt{2 \pi \hbar}$ and $E_p = p^2 / 2m \geq 0$ is the kinetic energy of a particle with mass $m$ \footnote{In general, the particle can also have internal degrees of freedom (e.g. spin). In that case, we would be concerned instead with the total (kinetic and internal) energy of the particle, adding complexity without changing the results of the study. For simplicity, we only describe the kinetic degrees of freedom of the particle and $\mathcal{H}_P$ is then the space of square-integrable functions on the real line.}. The interaction operator $\hat{V}(\hat{x})$ is assumed to vanish sufficiently far away from the scattering region $x \rightarrow \pm \infty$ \cite{Belkic2004,Taylor2006}. For such potentials, scattering theory guarantees the existence of a one-to-one operator between free (incoming) states and free (outgoing) states in $\mathcal{H}$. More specifically, $\mathcal{H}$ can be decomposed into the direct sum of scattering
and bound states and only the asymptotic behavior
of the former is that of free states, i.e., its evolution
for $t \rightarrow \pm \infty$ is given by the unitary evolution operator
$\hat{U}_0(t) = \exp(-i \hat{H}_0 t / \hbar)$ while the latter remain bound to the interaction
region in the same limit. If $\hat{U}(t) = \exp(-i\hat{H}t/\hbar)$ is the full
evolution operator, one can define the M{\o}ller operators
\begin{align}
    \hat{\Omega}_{\pm} \equiv \lim_{t \rightarrow \mp\infty} \hat{U}(t)^{\dagger}\hat{U}_0(t) 
\end{align}
which map incoming and outgoing states in $\mathcal{H}$ onto scattering states. The scattering operator defined as
\begin{align}
\label{scattop}
    \hat{S} \equiv \hat{\Omega}_{-}^{\dagger}\hat{\Omega}_{+} \; ,
\end{align}
is then unitary $\hat{S}^{\dagger} \hat{S} = \hat{S} \hat{S}^{\dagger} = \hat{\mathbb{I}}$ and satisfies energy conservation $[\hat{S},\hat{H}_0]=0$, mapping incoming states onto outgoing states. We note that one can prove that, under not too strict conditions, any state in $\cal H$ can be interpreted as a free (incoming or outgoing) state. In other words, the domain of the M{\o}ller and the scattering operator is the whole Hilbert space $\cal H$ \cite{Belkic2004,Taylor2006}. In this study, we are interested in incoming states where the particle is far from the collision region and approaching the system or scatterer either from the left or from the right. In the expressions presented above $\hat{\mathbb{I}}$, $\hat{\mathbb{I}}_S$ and $\hat{\mathbb{I}}_P$ denote the identity operators on $\mathcal{H}$, $\mathcal{H}_S$ and $\mathcal{H}_P$, respectively.

\subsection{Scattering amplitudes}

The scattering operator encodes the transition amplitudes for a collision. We define the kinetic energy eigenstates $\ket{E^{\alpha}_{p}} \equiv \sqrt{m/|p|} \ket{p}$, where $\alpha = \mathrm{sign}(p)$ accounts for the direction of the particle, which can be travelling to the left ($\alpha = +$) or right ($\alpha = -$). Then a full eigenbasis for $\hat{H}_0$ is $\{ \ket{E^{\alpha}_{p},j} \} \equiv \{ \ket{j} \otimes \ket{E^{\alpha}_{p}} \}$ and we omit the range of the index $j$ for simplicity. Due to the commutation relation $[\hat{S},\hat{H}_0]=0$, the scattering operator in this basis reads 
\begin{align}
    \label{scattopmatrix}
    \braket{E^{\alpha'}_{p'},j'|\hat{S}|E^{\alpha}_{p},j} = \delta(E_{p'} + e_{j'} - E_p - e_j) s^{\alpha' \alpha}_{j'j}(E_p + e_j) \; .
\end{align}
In the last expression, the $\delta$ function ensures energy conservation before and after the collision and $s_{j'j}^{\alpha' \alpha}(E)$ is the scattering matrix encoding the transition amplitudes from $\ket{E^{\alpha}_p,j} \rightarrow \ket{E^{\alpha'}_{p'},j'}$ at total energy $E = E_p + e_j$ \cite{Belkic2004,Taylor2006}. In practice, the scattering matrix can be calculated by solving the stationary
Schrödinger equation with appropriate asymptotic boundary
conditions \cite{Razavy2003,Belkic2004,Taylor2006,Jacob2021}. We show in Appendix~\ref{app:scattmatrixproperties} that the unitary property of $\hat{S}$ translates into the unitary property of the scattering matrix.

It is now useful to decompose the scattering operator as $\hat{S} = \hat{\mathbb{I}} - i \hat{T}$, where $\hat{T}$ is called the scattering amplitude operator. This decomposition is meaningful from a scattering perspective, since a collision always induces some change in the systems involved that is now captured by the operator $\hat{T}$. Substituting into Eq.~\eqref{scattopmatrix} we obtain the relation $\braket{E^{\alpha'}_{p'},j'|\hat{S}|E^{\alpha}_{p},j} = \delta_{\alpha' \alpha} \delta_{j'j}\delta(E_{p'} - E_p ) - i \braket{E^{\alpha'}_{p'},j'|\hat{T}|E^{\alpha}_{p},j}$
with
\begin{align}
    \label{Topmatrix}
    \braket{E^{\alpha'}_{p'},j'|\hat{T}|E^{\alpha}_{p},j} = \delta(E_{p'} + e_{j'} - E_p - e_j) t^{\alpha' \alpha}_{j'j}(E_p + e_j) \;.
\end{align}
In the last expression the scattering amplitude is related to the scattering matrix by through the expression 
\begin{align}
    \label{scattmatrix-decomposition}
    s^{\alpha' \alpha}_{j'j}(E_p + e_j) = \delta_{\alpha' \alpha}\delta_{j'j} -i t^{\alpha' \alpha}_{j'j}(E_p + e_j) \; .
\end{align}
We note that the unitary property of the scattering operator induces constraints on the operator $\hat{T}$. In particular, using $\hat{S} = \hat{\mathbb{I}} - i \hat{T}$ in the equality $\hat{S}^{\dagger} \hat{S} = \hat{\mathbb{I}}$ we arrive at $i(\hat{T} - \hat{T}^{\dagger}) = \hat{T}^{\dagger} \hat{T}$; proceeding the same way for $\hat{S}\hat{S}^{\dagger} = \hat{\mathbb{I}}$ we arrive at $i(\hat{T} - \hat{T}^{\dagger}) = \hat{T} \hat{T}^{\dagger}$. First, we conclude that $\hat{T}$ is a normal operator $[\hat{T},\hat{T}^{\dagger}]=0$. Second, the important relation $i(\hat{T} - \hat{T}^{\dagger}) = \hat{T}^{\dagger} \hat{T}$ imposes constraints on the imaginary part of the spectrum of $\hat{T}$. Decomposing $\hat{T}$ into its adjoint and anti-adjoint part we have $\hat{T} = \hat{T}_{\mathrm{H}} + i \hat{T}_{\mathrm{AH}}$, where $\hat{T}_{\mathrm{H}} = (\hat{T} + \hat{T}^{\dagger})/2$ is self-adjoint and $\hat{T}_{\mathrm{AH}} = (\hat{T} - \hat{T}^{\dagger})/2i = - \hat{T}^{\dagger} \hat{T} / 2$ is not only self-adjoint but a negative operator since $\hat{T}^{\dagger} \hat{T} \geq 0$. This leads to the so-called optical theorem of scattering theory as we show in Appendix~\ref{app:scattmatrixproperties}.

\subsection{The scattering map}

In this section, we derive the scattering dynamics on $\mathcal{H}_S$ starting from the full dynamics on $\mathcal{H}$. Let $\hat{\rho}$ be a density operator on $\mathcal{H}$ ($\hat{\rho} \geq 0$ and $\mathrm{Tr}[\hat{\rho}]=1$) representing the full quantum state before a scattering event. Then the quantum state after the scattering event $\hat{\rho}'$ is given by $\hat{\rho}' = \hat{S} \hat{\rho} \hat{S}^{\dagger}$ {and substituting the $\hat{T}$ decomposition of $\hat{S}$ we arrive at}
\begin{align}
    \hat{\rho}' & = \hat{S} \hat{\rho} \hat{S}^{\dagger} \nonumber \\ 
    & = \hat{\rho} + i \hat{\rho} \hat{T}^{\dagger} - i \hat{T}\hat{\rho} + \hat{T}\hat{\rho}\hat{T}^{\dagger} \nonumber \\
    \label{fullscattmap}
    & = \hat{\rho} - i [\hat{T}_{\mathrm{H}},\hat{\rho}] + \hat{T} \hat{\rho} \hat{T}^{\dagger} -\frac{1}{2} \{ \hat{T}^{\dagger} \hat{T}, \hat{\rho} \} \; .
\end{align}
In the last line, we substituted $\hat{T} = \hat{T}_{\mathrm{H}} + i \hat{T}_{\mathrm{AH}}$ in the terms with an prefactor of $i$ and $[\cdot,\cdot]$ and $\{ \cdot,\cdot \}$ denote the commutator and anti-commutator. Note that Eq.~\eqref{fullscattmap} still describes a unitary map. 

Assuming that the system and particle are uncorrelated before the collision $\hat{\rho} = \hat{\rho}_S \otimes \hat{\rho}_P$, we can take the partial trace $\mathrm{Tr}_P$ with respect to $\mathcal{H}_P$ in Eq.~\eqref{fullscattmap}. The final state of the system after the collision is then given by $\hat{\rho}_S' = \Phi(\hat{\rho}_S) \equiv \mathrm{Tr}_P [\hat{\rho}']$, where $\Phi$ is a CPTP map between density operators on $\mathcal{H}_S$ \cite{Alicki2007,Breuer2007,Rivas2012}. In order to take the partial trace in Eq.~\eqref{fullscattmap} explicitly, we need two ingredients. The first is the representation of $\hat{T}$ in the kinetic energy eigenbasis which reads 
\begin{align}
    \braket{E^{\alpha'}_{p'}|\hat{T}|E^{\alpha}_{p}} & = \sum_{j',j} \ket{j'}\bra{j} \braket{E^{\alpha'}_{p'},j'|\hat{T}|E^{\alpha}_{p},j} \nonumber \\
    & = \sum_{j',j} \ket{j'}\bra{j} \delta(E_{p'} + e_{j'} - E_p - e_j) t^{\alpha' \alpha}_{j'j}(E_p + e_j) \nonumber \\
    & = \sum_{\Delta} \delta(E_{p'} - E_p + \Delta) \hat{T}^{\alpha' \alpha}_{\Delta} (E_p) \; .
\end{align}
In the first line, we used twice the system's resolution of identity $\hat{\mathbb{I}}_S = \sum_{j} \ket{j}\bra{j}$ where we omit the summation limits for brevity. In the second line, we used Eq.~\eqref{Topmatrix}. In the third line, we rewrote the sum over $j',j$ as a sum over energy differences $\Delta$ where
\begin{align}
    \label{eigenoperators}
    \hat{T}^{\alpha' \alpha}_{\Delta}(E_p) \equiv \sum_{\substack{j',j:\\ e_{j'}-e_j = \Delta}} t_{j'j}^{\alpha' \alpha}(E_p + e_j) \ket{j'}\bra{j} \; ,
\end{align}
are eigenoperators of $\hat{H}_S$ and thus obey $[\hat{H}_S, \hat{T}^{\alpha' \alpha}_{\Delta}(E_p)] = \Delta \hat{T}^{\alpha' \alpha}_{\Delta}(E_p)$. The second ingredient is the representation of the particle's state in the same basis $\rho^{\alpha \alpha'}_P(E_{p},E_{p'}) \equiv \braket{E^{\alpha}_{p}|\hat{\rho}_P|E^{\alpha'}_{p'}}$. Using the particle's resolution of identity $\hat{\mathbb{I}}_P = \int dE_p \sum_{\alpha} \ket{E^{\alpha}_p} \bra{E^{\alpha}_p}$, where we also omit the integral and summation limits, we can now take the partial trace over the state of the particle in Eq.~\eqref{fullscattmap} and obtain the map for the system
\begin{align}
    \label{cptp}
    \hat{\rho}'_S = \hat{\rho}_S -i [\hat{H}_{LS},\hat{\rho}_S] + \mathcal{D}(\hat{\rho}_S) \; ,
\end{align}
where
\begin{align}
    \label{lambshift}
    \hat{H}_{LS} \equiv \frac{1}{2}\int dE_p \int dE_{p'}~\sum_{\Delta}\delta(E_{p'} - E_p + \Delta) ~\sum_{\alpha',\alpha}  \hat{T}_{\Delta}^{\alpha'\alpha}(E_p) \rho_P^{\alpha \alpha'}(E_p,E_{p'}) + \mathrm{c.c.} \; 
\end{align}
is the Lamb-Shift Hamiltonian, providing a {contribution to the dynamics that is first-order in the scattering amplitudes} and c.c. denotes the complex conjugate. The dissipator $\mathcal{D}$ is a superoperator {that is second-order in the amplitudes} 
\begin{align}
    \label{dissipator}
    \mathcal{D}(\hat{\rho}_S) & \equiv \int dE_p \int dE_{p'} \sum_{\Delta,\Delta'} \delta(\Delta - \Delta' - E_p + E_{p'})\sum_{\alpha',\alpha} \rho_P^{\alpha \alpha'}(E_p,E_{p'}) \nonumber \\
    & \times \sum_{\alpha''} \big[ \hat{T}_{\Delta}^{\alpha''\alpha}(E_p) \hat{\rho}_S \hat{T}_{\Delta'}^{\alpha''\alpha'}(E_{p'})^{\dagger} - \frac{1}{2} \{ \hat{T}_{\Delta'}^{\alpha''\alpha'}(E_{p'})^{\dagger} \hat{T}_{\Delta}^{\alpha''\alpha}(E_p) , \hat{\rho}_S \}  \big] \; .
\end{align}
{The terms \textit{Lamb-Shift Hamiltonian} and \textit{dissipator} are used here in analogy with the structure of the more familiar quantum master equations \cite{Alicki2007,Breuer2007,Rivas2012}; we expand on this issue below.} 

If $\hat{A}$ is a self-adjoint operator on $\mathcal{H}_S$ representing a system observable, then its change as a result of the collision is given by $\delta A = \mathrm{Tr}_S[\hat{A} \hat{\rho}'_S] - \mathrm{Tr}_S[\hat{A} \hat{\rho}_S]$ where $\mathrm{Tr}_S$ is the trace over $\mathcal{H}_S$. Using Eq.~\eqref{cptp} we can write $\delta A = \delta A_{LS} + \delta A_{\mathcal{D}}$, where
\begin{align}
    \label{changelambshift}
    \delta A_{LS} & \equiv -i \mathrm{Tr}_S [[\hat{A},\hat{H}_{LS}]\hat{\rho}_S] \; , \\
    \label{changedissipator} \delta A_{\mathcal{D}} & \equiv \mathrm{Tr}_S[\hat{A} \mathcal{D}(\hat{\rho}_S)] \; ,
\end{align}
and we used the cyclic property of the trace to achieve Eq.~\eqref{changelambshift}.

Let us now discuss the map defined by Eqs.~\eqref{cptp}, \eqref{lambshift} and \eqref{dissipator}. {We start by noting that the structure of the map in Eq.~\eqref{cptp} bears similarities with the structure of quantum master equations \cite{Alicki2007,Breuer2007,Rivas2012}. Nevertheless, Eq.~\eqref{cptp} is an exact map describing the dynamics of a single collision event, as opposed to the approximate continuous-time descriptions using master equations. In order to make this connection clearer, we derive in Appendix~\ref{app:qme} a quantum master equation for the average state of the system after repeated collisions with the particle, assuming that the waiting times between collisions are Poisson-distributed with a rate $\gamma$, i.e. with average waiting time $\gamma^{-1}$. The quantum master equation reads
\begin{align}
    \label{qme}
    \frac{d  \hat{\bar\rho}_S(t)}{dt} = -\frac{i}{\hbar} [\hat{H}_S + \hbar\gamma \hat{H}_{LS}, \hat{\bar\rho}_S(t)] + \gamma \mathcal{D}(\hat{\bar\rho}_S(t)) \; .
\end{align}
where $\hat{\bar{\rho}}_S(t)$ denotes the average state of the system at time $t$. From Eq.~\eqref{qme}, we see that while $\mathcal{D}$ describes the dissipative part of the dynamics, the unitary part is governed by the effective Hamiltonian $\hat{H}_{\mathrm{eff}} \equiv \hat{H}_S + \hbar \gamma \hat{H}_{LS}$. Despite Eq.~\eqref{qme} having the familiar form of a quantum master equation, the precise form of Eqs.~\eqref{lambshift} and \eqref{dissipator} are still different from those derived in the Born-Markov-Secular approximation or scattering of gas particles in the low-density limit \cite{Alicki2007,Breuer2007,Rivas2012,Dumcke1985}. Namely, the Lamb-Shift Hamiltonian does not commute with the system Hamiltonian $[\hat{H}_{LS},\hat{H}_S] \neq 0$, while the dissipator in Eq.~\eqref{dissipator} is not in usual Lindblad form since the system eigenoperators occur with different energy differences $\Delta \neq \Delta'$. 

The essential feature of Eqs.~\eqref{lambshift} and \eqref{dissipator} -- which define the map in Eq.~\eqref{cptp} and, consequently, the quantum master equation in Eq.~\eqref{qme} -- is that they depend} crucially on the quantum state of the particle with respect to the energy differences of the system. Due to the $\delta$ functions in Eqs.~\eqref{dissipator} or \eqref{lambshift}, the particle's state reads $\rho_P^{\alpha \alpha'}(E_p,E_{p} + \Delta' - \Delta)$ and thus depends on superpositions (coherences) in the basis of kinetic energy shifted by the energy differences of the system. In particular, if the particle's state is sufficiently narrow with respect to the system's energy differences (and if it is travelling with a well-defined direction $\alpha$) then $\rho_P^{\alpha \alpha'}(E_p,E_{p} + \Delta' - \Delta) \simeq \delta_{\alpha \alpha'} \delta_{\Delta \Delta'} \rho_P^{\alpha}(E_p)$ where $\rho_P^{\alpha}(E_p) \equiv \rho_P^{\alpha}(E_p,E_{p})$ is the kinetic energy distribution of the particle. These narrow states describe particles which are well-localized in energy but delocalized in space, and thus can be well approximated by a plane wave or a mixture of plane waves \cite{Jacob2021}. Taking these states and performing the integrals in Eq.~\eqref{lambshift} and \eqref{dissipator} leads to
\begin{gather}
    \label{lambshift-narrow}
    \hat{H}_{LS} = \frac{1}{2}\int dE_p~\sum_{\alpha}  \hat{T}_{0}^{\alpha\alpha}(E_p) \rho_P^{\alpha}(E_p) + \mathrm{c.c.} \; , \\
    \label{dissipator-narrow}
    \mathcal{D}(\hat{\rho}_S) = \int dE_p \sum_{\alpha} \rho_P^{\alpha}(E_p) \sum_{\alpha'' \Delta} \big[ \hat{T}_{\Delta}^{\alpha''\alpha}(E_p) \hat{\rho}_S \hat{T}_{\Delta}^{\alpha''\alpha}(E_{p})^{\dagger} - \frac{1}{2} \{ \hat{T}_{\Delta}^{\alpha''\alpha}(E_{p})^{\dagger} \hat{T}_{\Delta}^{\alpha''\alpha}(E_p) , \hat{\rho}_S \}  \big] \; .
\end{gather}
In this case $[\hat{H}_S,\hat{T}_{0}^{\alpha\alpha}(E_p)] = [\hat{H}_S,\hat{H}_{LS}] = 0$ and the dissipator is in Lindblad form, in line with well-known microscopic derivations of quantum master equations with incoherent -- usually thermal -- {baths  \cite{Spohn1978,Dumcke1985,Breuer2007,Alicki2007,Rivas2012}}; in particular, Eqs.~\eqref{lambshift-narrow} and \eqref{dissipator-narrow} have the same form {(up to multiplicative constant)} as the Lindblad generator derived in the low-density limit \cite{Dumcke1985}. Since now the Lamb-Shift only provides a simple renormalization of system Hamiltonian, it is usually ignored in many applications of quantum master equations and the dissipator plays the major role in the dynamics and changes in observables. An example are Refs.~\cite{Jacob2021,Tabanera2022,Jacob2023,Tabanera2023,Jacob2024}, where Eqs.~\eqref{lambshift-narrow} and \eqref{dissipator-narrow} were implicitly used (although not explicitly derived) to study {thermalisation and }heat exchanges \cite{Jacob2021,Tabanera2022,Tabanera2023}, as well as energy fluctuations in scattering theory \cite{Jacob2023,Jacob2024}. {In these cases, the relevant observable is energy} $\hat{A} = \hat{H}_S$ {and $\hat{H}_{LS}$} defined by Eq.~\eqref{lambshift-narrow} plays no role in energy changes -- Eq.~\eqref{changelambshift} vanishes and all energy change is dissipative and given by Eq.~\eqref{changedissipator}. {Although narrow states guarantee that the populations of the system decouple from coherences under quite general conditions \cite{Jacob2021,Tabanera2022,Tabanera2023}, more ingredients are necessary to achieve thermalisation under repeated collisions, i.e. under repeated applications of the map in Eq.~\eqref{cptp}: the collision has to be microscopically reversible and the kinetic energy of the particle has to be thermally distributed $\rho^{\alpha}_P(E_p) \sim e^{-\beta E_p}$, where $\beta \equiv 1/(k_B T)$, $k_B$ is Boltzmann's constant and $T$ is the temperature. One can then show that detailed-balance holds and that $\hat{\omega}_{\beta} \equiv e^{-\beta \hat{H}_S} / Z$ is the fixed point of the map in Eq.~\eqref{cptp} (and consequently of the quantum master equation in Eq.~\eqref{qme}), where $Z = \mathrm{Tr}_S[e^{-\beta \hat{H}_S}]$ is the partition function. \cite{Jacob2021,Tabanera2022,Tabanera2023}.} 

The situation is different if the particle states are not narrow with respect to the system energy differences, in which case $[\hat{H}_S,\hat{H}_{LS}] \neq 0$ and the {Lamb-Shift Hamiltonian} defined by Eq.~\eqref{lambshift} might play a major role in the dynamics, energy changes and changes in other observables. Evidence for this comes from the opposite limiting case, where states are very broad with respect to the energy differences, i.e. represented by spatially localized states instead of plane waves. In scattering theory, very massive and localized particles were shown to mimick a time-dependent interaction and induce unitary (instead of dissipative) evolution on the system, thus playing the role of a work source \cite{Jacob2022,Piccione2024}. This raises the question of whether linear response theory, which relies on a time-dependent perturbation and unitary evolution, can also be formulated within the time-independent framework of scattering theory. {Indeed, we start the next section by showing that a non-perturbative response, satisfying a certain fluctuation-dissipation relation, arises at the level of the Lamb-Shift Hamiltonian in Eq.~\eqref{lambshift}. We expect \textit{a priori} that linear response theory, if it can be formulated within scattering theory under some conditions, should emerge from the Lamb-Shift Hamiltonian rather than the dissipator. This is because the change in an observable $\hat{A}$ caused by this Hamiltonian, given by Eq.~\eqref{changelambshift}, has the same structure as that of linear response theory -- it involves a commutator of $\hat{A}$ with an operator that is first-order in the scattering amplitudes $\hat{T}$. By the end of the section, we show how Kubo's formula arises from the Born approximation, i.e. in the weak coupling regime of scattering theory, when the contribution of the dissipator can be ignored.}

\section{Response of the quantum system to a collision \label{sec:response}}

\subsection{Non-perturbative response}

The Lamb-Shift perturbation $\hat{H}_{LS}$ in Eq.~\eqref{lambshift} defines {the scattering dynamics at first-order in the scattering amplitudes}, inducing a change in observable $\hat{A}$ given by Eq.~\eqref{changelambshift}. Exploring the cyclic property of the trace and since $\hat{H}_{LS}$ is self-adjoint, we can write $\delta A_{LS} = \Re[\chi_A]$, where $\Re$ is the real part and $\chi_A$ is a complex number given by
\begin{align}
    \label{chiA}
    \chi_{A} \equiv \int dE_p \int dE_{p'}~\sum_{\Delta}\delta(E_{p'} - E_p + \Delta) ~\sum_{\alpha',\alpha}  \chi_{\Delta}^{\alpha'\alpha}(E_p) \rho_P^{\alpha \alpha'}(E_p,E_{p'}) \; .
\end{align}
In the last expression, the quantity of interest is given by
\begin{align}
    \label{response-Delta}
    \chi_{\Delta}^{\alpha'\alpha}(E_p) \equiv -i \mathrm{Tr}_S[[\hat{A},\hat{T}_{\Delta}^{\alpha'\alpha}(E_p)]\hat{\rho}_S] \; ,
\end{align}
which we claim encodes the non-perturbative response of observable $\hat{A}$ to a quantum jump of energy $\Delta$ induced by a collision with the particle having initial kinetic energy $E_p$ and direction $\alpha$, and final kinetic energy $E_p - \Delta$ and direction $\alpha'$. By non-perturbative, we mean that $\hat{T}_{\Delta}^{\alpha'\alpha}(E_p)$ encodes the exact scattering amplitudes for a collision. {We now show that Eq.~\eqref{response-Delta} can be understood as the discrete Fourier transform of a response function, encoding the change in observable $\hat{A}$ induced the collision.} Making use of the representation of the $\delta$ function
\begin{align}
    \label{deltafunction}
    \delta(E_{p'} - E_p + \Delta) = \frac{1}{2 \pi \hbar} \int^{+\infty}_{-\infty} dt~ e^{i(\Delta - E_p + E_{p'})t/\hbar} \; ,
\end{align}
which we substitute in Eq.~\eqref{chiA}. We can then perform the following sum over energy differences
\begin{align}
    \sum_{\Delta} e^{i \Delta t /\hbar} \hat{T}_{\Delta}^{\alpha'\alpha}(E_p) = e^{i \hat{H}_S t /\hbar} \sum_{\Delta} \hat{T}_{\Delta}^{\alpha'\alpha}(E_p) e^{-i \hat{H}_S t /\hbar} = e^{i \hat{H}_S t /\hbar} \hat{\mathcal{T}}^{\alpha'\alpha}(E_p) e^{-i \hat{H}_S t /\hbar} =  \hat{\mathcal{T}}^{\alpha'\alpha}(E_p,t) \nonumber \; ,
\end{align}
where $\hat{O}(t) \equiv e^{i \hat{H}_S t/\hbar} \hat{O} e^{-i \hat{H}_S t/\hbar}$ denotes the Heisenberg picture with respect to $\hat{H}_S$ for an observable $\hat{O}$ on $\mathcal{H}_S$, we used the properties of the eigenoperators in Eq.~\eqref{eigenoperators} and defined 
\begin{align}
    \label{amplitude-operator}
    \hat{\mathcal{T}}^{\alpha' \alpha}(E_p) \equiv \sum_{j',j} t_{j'j}^{\alpha' \alpha}(E_p + e_j) \ket{j'}\bra{j} \; .
\end{align}
Moreover, we recognize that 
\begin{align}
    \rho_P^{\alpha \alpha'}(E_p,E_{p'}) e^{-i E_p t /\hbar} e^{i E_{p'} t /\hbar} = \braket{E^{\alpha}_{p}|\hat{\rho}_P|E^{\alpha'}_{p'}} e^{-i E_p t /\hbar} e^{i E_{p'} t /\hbar} = \braket{E^{\alpha}_{p}|\hat{\rho}_P(t)|E^{\alpha'}_{p'}} \equiv \rho_P^{\alpha \alpha'}(E_p,E_{p'},t) \nonumber \; ,
\end{align}
where $\rho_P(t) \equiv e^{-it \hat{p}^2 / 2m\hbar}\hat{\rho}_P e^{it \hat{p}^2 / 2m\hbar}$ is the evolved particle state. After these considerations, Eq.~\eqref{chiA} becomes
\begin{align}
    \label{chiA-2}
    \chi_{A} = \int^{+\infty}_{-\infty} dt \int dE_p \int dE_{p'} 
 \sum_{\alpha',\alpha}~\chi_A^{\alpha' \alpha}(E_p,-t) \rho_P^{\alpha \alpha'}(E_p,E_{p'},t) \; ,
\end{align}
where
\begin{align}
    \label{response-time}
    \chi_A^{\alpha' \alpha}(E_p,t) \equiv \frac{1}{2 \pi \hbar}\sum_{\Delta} e^{-i \Delta t / \hbar} \chi^{\alpha' \alpha}_{\Delta}(E_p) = -\frac{i}{2 \pi \hbar} \mathrm{Tr}_S[[\hat{A},\hat{\mathcal{T}}^{\alpha'\alpha}(E_p,-t)] \hat{\rho}_S] \; .
\end{align}
Eqs.~\eqref{chiA-2} and \eqref{response-time} constitute our first result. Eq.~\eqref{response-time} is the equivalent of Eq.~\eqref{response-Delta} now expressed in the time domain in the form of two-time correlation between the operators $\hat{\mathcal{T}}^{\alpha'\alpha}(E_p)$ and $\hat{A}$. More concretely $\chi^{\alpha' \alpha}_{\Delta}(E_p)$ is the discrete Fourier transform of $\chi_A^{\alpha' \alpha}(E_p,t)$; an expression in terms of the more familiar continuous Fourier transform is also possible as we show in Appendix~\ref{app:fouriertransform}. The response is then integrated over time and over the full state of the particle in Eq.~\eqref{chiA-2}, where the latter plays a role similar to a time-dependent force function. It is important to note that, in opposition to linear response theory for closed systems, the perturbation $\hat{\mathcal{T}}^{\alpha'\alpha}(E_p)$ defined in Eq.~\eqref{amplitude-operator} is not self-adjoint \footnote{The fact that the amplitude operator $\hat{T}$ and, by consequence, $\hat{\mathcal{T}}^{\alpha'\alpha}(E_p)$ are not self-adjoint has other implications already pointed out in previous works. For example, it can lead to violations of detailed balance in quantum master equations in the low-density limit \cite{Alicki2023}} and captures the effect of the collision on the system in a non-perturbative fashion. As a result, the non-perturbative response function $\chi_A^{\alpha' \alpha}(E_p,t)$ is generally a complex quantity. We show later that, in the Born approximation, this quantity encodes Kubo's formula in terms of a real response function.

We now 
consider that $\hat{\rho}_S = \hat{\omega}_{\beta}$ where is the thermal state $\hat{\omega}_{\beta} \equiv e^{-\beta \hat{H}_S} / Z$, $Z = \mathrm{Tr}_S[e^{-\beta \hat{H}_S}]$ is the partition function and $\beta \equiv 1/(k_B T)$ where $k_B$ is Boltzmann's constant and $T$ is the temperature. Using the cyclic property of the trace, it is easy to show that the response function is homogeneous in time $\mathrm{Tr}_S[[\hat{A}(0),\hat{\mathcal{T}}^{\alpha'\alpha}(E_p,-t)] \hat{\omega}_{\beta}] = \mathrm{Tr}_S[[\hat{A}(t_0),\hat{\mathcal{T}}^{\alpha'\alpha}(E_p,t_0 -t)] \hat{\omega}_{\beta}]$ for any real $t_0$. Then we can write Eq.~\eqref{response-time} in a more recognizable form
\begin{align}
    \label{response-time-thermal}
    \chi_A^{\alpha' \alpha}(E_p,t) = -\frac{i}{2 \pi \hbar} \mathrm{Tr}_S[[\hat{A}(t),\hat{\mathcal{T}}^{\alpha'\alpha}(E_p)] \hat{\omega}_{\beta}] \; ,
\end{align}
expressing the response of an observable $\hat{A}$ at time $t$ to the perturbation $\hat{\mathcal{T}}^{\alpha'\alpha}(E_p)$ at time $t=0$. Note that, in linear response theory for closed systems, causality is imposed by the fact that Eq.~\eqref{response-time-thermal} is only taken for $t > 0$; or equivalently, that the integral in time is a convolution between the force function at an earlier time and the response at a later time (see Appendix \ref{app:linearresponse}). Instead, in our autonomous scattering framework, no such condition is generally imposed by the theory \footnote{This is similar to what happens in the traditional neutron scattering formalism, where certain scattering functions encoding correlations are defined for all times \cite{Scherm1972,Berk1993,Irfan2021}}. This notion of causality is recovered later when we take the Born approximation together with spatially localized particle states.

\subsection{Fluctuation-dissipation relation}

Fluctuation-dissipation relations connect the Fourier transform of a response function to the Fourier transform of correlations \cite{Kubo1957-I,Kubo1966}. We prove that a similar relation also holds for the non-perturbative response in Eq.~\eqref{response-time-thermal}. The discrete Fourier transform $\chi_{\Delta}^{\alpha'\alpha}(E_p)$ in Eq.~\eqref{response-Delta} obeys the following relation for a thermal state
\begin{align}
    & \mathrm{Tr}_S[[\hat{A},\hat{T}_{\Delta}^{\alpha'\alpha}(E_p)]\hat{\omega}_{\beta}]  = \mathrm{Tr}_S[\hat{A}[\hat{T}_{\Delta}^{\alpha'\alpha}(E_p),\hat{\omega}_{\beta}]] \nonumber \\
    =~& (1 - e^{-\beta \Delta}) \mathrm{Tr}_S[\hat{A}\hat{T}_{\Delta}^{\alpha'\alpha}(E_p) \hat{\omega}_{\beta}] = \frac{1 - e^{-\beta \Delta}}{1 + e^{-\beta \Delta}} \mathrm{Tr}_S[\hat{A}\{\hat{T}_{\Delta}^{\alpha'\alpha}(E_p), \hat{\omega}_{\beta}\}] \nonumber \\
    =~& \tanh\bigg(\frac{\beta \Delta}{2}\bigg) \mathrm{Tr}_S[\{ \hat{A},\hat{T}_{\Delta}^{\alpha'\alpha}(E_p)\} \hat{\omega}_{\beta}] \nonumber \; .
\end{align}
In the first line, we used the cyclic property of the trace. In the second line we used the easily provable properties of the eigenoperators $\hat{\omega}_{\beta} \hat{T}_{\Delta}^{\alpha'\alpha}(E_p) = e^{-\beta \Delta} \hat{T}_{\Delta}^{\alpha'\alpha}(E_p) \hat{\omega}_{\beta}$. In the third line we recognize the hyperbolic tangent function and used the cyclic property of the trace again. If we now define
\begin{align}
    C_{\Delta}^{\alpha'\alpha}(E_p) \equiv \frac{1}{2}\mathrm{Tr}_S[\{ \hat{A},\hat{T}_{\Delta}^{\alpha'\alpha}(E_p)\} \hat{\omega}_{\beta}]  \; 
\end{align}
as the Fourier transform of the correlation between observable $\hat{A}$ and the perturbation $\hat{T}_{\Delta}^{\alpha'\alpha}(E_p)$ we arrive at the following relation
\begin{align}
    \label{fdr-Delta}
    \chi_{\Delta}^{\alpha'\alpha}(E_p) = - 2i \tanh\bigg(\frac{\beta \Delta}{2}\bigg) C_{\Delta}^{\alpha'\alpha}(E_p) \; .
\end{align}
The last expression constitutes our second result, showing a relationship between the Fourier transform of the response function in Eq.~\eqref{response-time-thermal} and the Fourier transform of the two-time correlation
\begin{align}
    \label{correlation-time}
    C_A^{\alpha' \alpha}(E_p,t) \equiv \frac{1}{2 \pi}\sum_{\Delta} e^{-i \Delta t / \hbar} C^{\alpha' \alpha}_{\Delta}(E_p) = \frac{1}{4 \pi} \mathrm{Tr}_S[\{\hat{A}(t),\hat{\mathcal{T}}^{\alpha'\alpha}(E_p)\} \hat{\omega}_{\beta}] \; ,
\end{align}
which, in analogy with Eq.~\eqref{response-time-thermal}, is homogeneous in time and complex due to the fact that $\hat{\mathcal{T}}^{\alpha'\alpha}(E_p)$ is not self-adjoint. Eq.~\eqref{fdr-Delta} implies for the imaginary $\Im$ and real $\Re$ part of the discrete Fourier transform of the response that
\begin{align}
    \label{fdr-Delta-Im}
    \Im[\chi_{\Delta}^{\alpha'\alpha}(E_p)] & = - 2\tanh\bigg(\frac{\beta \Delta}{2}\bigg) \Re[C_{\Delta}^{\alpha'\alpha}(E_p)] \; , \\
    \label{fdr-Delta-Re}
    \Re[\chi_{\Delta}^{\alpha'\alpha}(E_p)] & =  2\tanh\bigg(\frac{\beta \Delta}{2}\bigg) \Im[C_{\Delta}^{\alpha'\alpha}(E_p)] \; .
\end{align}
In Appendix~\ref{app:fouriertransform}, we show how an analogous relation to Eq.~\eqref{fdr-Delta} holds for the continuous Fourier transform. We note that non-perturbative nature of Eqs.~\eqref{response-time-thermal} and \eqref{correlation-time} does not allow us to express these quantities in terms of auto-correlation functions: while $\hat{A}$ is a self-adjoint operator and represents an observable, $\hat{\mathcal{T}}^{\alpha'\alpha}(E_p)$ is not self-adjoint and represents the full effect of the collision. This means that Eq.~\eqref{fdr-Delta} cannot be reduced to the usual fluctuation-dissipation relation for auto-correlations (see Appendix~\ref{app:linearresponse}). 

{Finally, we note that our results -- given by Eqs.~\eqref{chiA-2}, \eqref{response-time} and \eqref{fdr-Delta} -- show that the same mathematical structure of linear response theory and fluctuation-dissipation relations can be defined non-perturbatively and not only near thermal equilibrium. We show below how to recover linear response in the Born approximation, when the scattering interaction is weak.}

\subsection{Born approximation}

In scattering theory, the Born approximation consists in approximating the scattering amplitudes by the scattering potential
\begin{align}
    \label{born}
    t_{j'j}^{\alpha' \alpha}(E_p + e_j) \simeq 2 \pi \braket{E^{\alpha'}_{p} + e_j - e_{j'},j'|\hat{V}(\hat{x})|E^{\alpha}_{p},j} \; .
\end{align}
Although the validity of the Born approximation depends on the specific potential in use, it is generally valid for sufficiently weak potentials compared to the other terms in the Hamiltonian or at sufficiently high kinetic energies \cite{Belkic2004,Taylor2006}. Moreover, we can generally write the potential as $\hat{V}(\hat{x}) = \sum_{l} \hat{V}^{l}_S \otimes \hat{V}^{l}_P(\hat{x})$ where $l$ is an arbitrary index and $\hat{V}^{l}_S$ and $\hat{V}^{l}_P(\hat{x})$ are self-adjoint operators on $\mathcal{H}_S$ and $\mathcal{H}_P$, respectively \footnote{For example, the index $l$ can label scattering centers $\hat{V}^{l}_P(\hat{x}) = \lambda_l \delta(\hat{x}-x_l)$ {where $x_l$ are their positions and $\lambda_l$ coupling parameters}}. Substituting the Born approximation in Eq.~\eqref{born} into Eq.~\eqref{eigenoperators} leads to
\begin{align}
    \label{eigenoperators-Born}
    \hat{T}^{\alpha' \alpha}_{\Delta}(E_p) = 2\pi \sum_l \braket{E^{\alpha'}_{p} -\Delta|\hat{V}^{l}_P(\hat{x})|E^{\alpha}_{p}} \hat{V}^{l}_{\Delta} \; ,
\end{align}
where we defined the new eigenoperators in the Born approximation
\begin{align}
    \label{eigenoperators-real}
    \hat{V}^{l}_{\Delta} \equiv \sum_{\substack{j',j:\\ e_{j'}-e_j = \Delta}} \braket{j'|\hat{V}^{l}_S|j}\ket{j'}\bra{j} \; .
\end{align}
Analogously, we find that the response defined in Eq.~\eqref{response-Delta} also factorizes in the Born approximation
\begin{align}
     \chi_{\Delta}^{\alpha'\alpha}(E_p) = 2\pi \sum_l \braket{E^{\alpha'}_{p} -\Delta|\hat{V}^{l}_P(\hat{x})|E^{\alpha}_{p}} \chi^{l}_{\Delta} \; ,
\end{align}
where we identified the discrete Fourier transform of the response
\begin{align}
    \chi^{l}_{\Delta} \equiv -i \mathrm{Tr}_S[[\hat{A},\hat{V}_{\Delta}]\hat{\rho}_S] \; .
\end{align}
We then obtain Eq.~\eqref{chiA} in the Born approximation
\begin{align}
    \label{chiA-Born}
    \chi_{A} = 2 \pi \sum_l \int dE_p \int dE_{p'}~\sum_{\Delta}\delta(E_{p'} - E_p + \Delta)~\chi^{l}_{\Delta}~\sum_{\alpha',\alpha} \braket{E^{\alpha'}_{p'}|\hat{V}^{l}_P(\hat{x})|E^{\alpha}_{p}} \rho_P^{\alpha \alpha'}(E_p,E_{p'}) \; .
\end{align}
We can perform the same calculations as we did after Eq.~\eqref{response-Delta} and until Eq.~\eqref{chiA-2}. We now find that the integrals over the particle state in Eq.~\eqref{chiA-Born} can be written as a trace over $\mathcal{H}_P$, so that Eq.~\eqref{chiA-2} has a simple form in the Born approximation
\begin{align}
    \label{chiA-2-Born}
    \chi_{A} = \sum_l \int^{+\infty}_{-\infty} dt~\chi^{l}_A(-t) f^{l}(t) \; .
\end{align}
In the last expression, we have defined the response function in the Born approximation
\begin{align}
    \label{response-time-Born}
    \chi^{l}_A(t) & \equiv \frac{1}{\hbar}\sum_{\Delta} e^{-i \Delta t / \hbar} \chi^{l}_{\Delta} = -\frac{i}{ \hbar} \mathrm{Tr}_S[[\hat{A},\hat{V}_S^{l}(-t)] \hat{\rho}_S] \; ,
\end{align}
and the force function
\begin{align}
    \label{force}
    f^{l}(t) & \equiv \mathrm{Tr}_P[\hat{V}^{l}_P(\hat{x}) \hat{\rho}_P(t)] \; .
\end{align}
Eq.~\eqref{chiA-2-Born}, with definitions \eqref{response-time-Born} and \eqref{force}, constitute our third result. In opposition to Eq.~\eqref{response-time}, the perturbation in Eq.~\eqref{response-time-Born} is now given by a self-adjoint operator and the response function is now real. Since Eq.~\eqref{force} is also real, we conclude that Eq.~\eqref{chiA-2-Born} is real and thus $\chi_A = \delta A_{LS}$. When the initial state is thermal, the response function is homogeneous in time $\chi^{l}_A(t) = (1/i\hbar) \mathrm{Tr}_S[[\hat{A}(t),\hat{V}_S^{l}(0)] \hat{\omega}_\beta]$; and the equivalent of relation Eq.~\eqref{fdr-Delta} holds for $\chi^{l}_{\Delta}$ and $C^{l}_{\Delta} \equiv \mathrm{Tr}_S[\{ \hat{A},\hat{V}^l_{\Delta}\} \hat{\omega}_{\beta}]/2$ where the latter is the discrete Fourier transform of the real correlation $C^{l}_A(t) \equiv \mathrm{Tr}_S[\{\hat{A}(t),\hat{V}^l_{S}\} \hat{\omega}_{\beta}]/2$. In addition, the function in Eq.~\eqref{force} plays the role of a time-dependent force in the usual Kubo formalism, where the time-dependence comes from the particle's evolution through the scattering potential in real space.

In the Born approximation, the response of observable $\hat{A}$ in Eqs.~\eqref{changelambshift} or \eqref{response-time-Born} becomes first-order in the interaction potential, while the response induced by dissipative effects in Eq.~\eqref{changedissipator} becomes second-order. Therefore, insofar as the dissipative part can be ignored with respect to {Lamb-Shift Hamiltonian}, we can consider Eq.~\eqref{response-time-Born} as the complete response $\chi_A = \delta A_{LS} \simeq \delta A$. Whether this happens in practice depends not only on the state of the particle as we previously discussed, but also on the specific observable $\hat{A}$ and on the initial state of the system.

\subsection{Kubo's formula}

\begin{figure}[t!]
\centering
\includegraphics[width=0.5\textwidth]{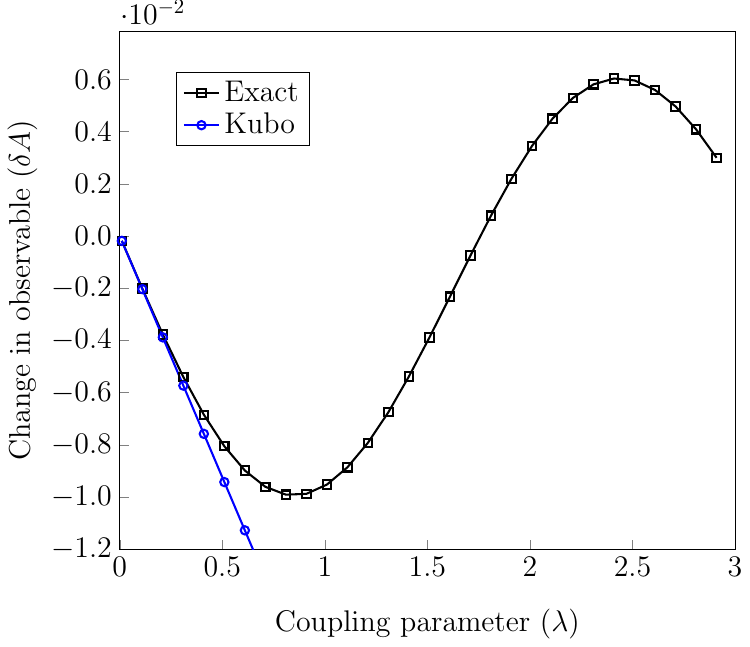}
\caption{\label{fig} {Change in observable $\delta A$ -- computed from exact scattering dynamics using $\delta A = \delta A_{LS} + \delta A_{\mathcal{D}}$ with Eqs.~\eqref{changelambshift} and \eqref{changedissipator}, and from Kubo's formula given by Eq.~\eqref{kubo} -- as a function of the coupling parameter $\lambda$. For small $\lambda$, where the Born approximation is valid, Kubo's formula matches the exact scattering dynamics. Regarding the details of the model, we chose a two-level system with energy gap $\Delta$ defined by the Hamiltonian $\hat{H}_S = \Delta \hat{\sigma}_{z}/2$ and a potential barrier of strength $V_0$ and length $a$ given by $\hat{V}(\hat{x}) = V_0[\hat{\sigma}_x \otimes\Theta(\hat{x}+a/2)\Theta(a/2-\hat{x})]$. We chose $\hat{A} = \hat{\sigma}_x$ and the coupling parameter $\lambda = V_0 a / \hbar v_0 $ \cite{Jacob2022} was varied by changing only $V_0$. For the exact scattering dynamics, we first computed the scattering matrix by solving numerically the multichannel scattering equations \cite{Razavy2003} and then considered the initial state of the particle to be a pure Gaussian state defined by the average momentum $p_0 = m v_0$ and position $x_0$, as well as $\sigma_x \sigma_p = \hbar/2$ with $\sigma_p$ and $\sigma_x$ being the uncertainty in momentum and position, respectively. In order for the particle to be travelling semiclassicaly and sufficiently localized, the conditions $E_{p_0} \gg V_0$, $p_0 a \gg \hbar$ must hold, together with the inequality $p_0 \gg \sigma_p \gg \Delta /v_0$ (see Ref.~\cite{Jacob2022} for more details). The initial state of the system was chosen to be thermal $\omega_{\beta}$. Numerical values: $\hbar = \Delta = m = a = \beta = 1$, $x_0 = 2$, $p_0 = v_0 = 100$, $\sigma_p = 0.2$. Note that, although we have considered throughout our study that the system $S$ is fixed and the particle $P$ is moving, this model can also be interpreted as a two-level atom $A = S + P$ flying through a spin-dependent potential; Eq.~\eqref{cptp} then describes the dynamics of the internal state of the atom \cite{Piccione2024}.}}
\end{figure}

Although Eq.~\eqref{chiA-2-Born} is very similar to Kubo's formula, it still differs in the fact that the integral is not a convolution of the force and response functions as in linear response theory for closed systems (see Appendix \ref{app:linearresponse}). To make progress, we compute Eq.~\eqref{force} in the position $\{ \ket{x} \}$ eigenbasis of $\mathcal{H}_P$ and assume an incoming particle from the left \footnote{This is without loss of generality and we could have equally considered a particle approaching from the right.} that is very massive and localized in space with respect to the potential in space. This approximation means that the particle's motion can be taken as classical in the sense that $\braket{x|\hat{\rho}_P(t)|x} = \delta (x - x_0 - v_0 t)$ where $x_0$ and $v_0 \geq 0$ are the average position and velocity at $t=0$, respectively (see e.g. Ref.~\cite{Jacob2022,Piccione2024} for interesting applications of massive and localized particles). Eq.~\eqref{force} reads
\begin{align}
    \label{force-delta}
    f^{l}(t) = \int dx~V^{l}(x) \braket{x|\hat{\rho}_P(t)|x} = V^{l}(x_0 + v_0 t) \; ,
\end{align}
where $V^{l}(x) \equiv \braket{x|\hat{V}^{l}_P|x}$. Let us now assume that the potential vanishes after some point in space, i.e. $V^{l}(x) = 0$ for some $x \geq x_{\mathrm{max}}$ \footnote{To be more precise we require $V^{l}(x) = 0$ for all $l$ which always happens for $x \geq \max_{l} (\{ x^{l}_{\mathrm{max}} \})$}. Since $x_0$ is arbitrary, in the sense that the integral in Eq.~\eqref{chiA-2-Born} does not change its value for different $x_0$, we choose it to lie at the border or beyond the potential region $x_0 \geq x_{\mathrm{max}}$. The integral in Eq.~\eqref{chiA-2-Born} can then be written $\chi_{A} = \sum_l \int^{0}_{-\infty} dt'~\chi^{l}_A(-t') V_P^{l}(x_0 + v_0 t')$ where we restricted the upper limit to zero without detriment. We can now shift the origin of time through a change of variables $\tau \equiv t + t'$, where $t \geq \tau$ since $t'$ is negative in the integrand, and define $\lambda^l(\tau) \equiv V^{l}_P(x_0 - v_0 (t - \tau))$ as a function of time, arriving at
\begin{align}
    \label{kubo}
    \chi_{A} = \sum_l \int^{t}_{-\infty} d\tau~\chi^{l}_A(t-\tau) \lambda^{l}(\tau) \; .
\end{align}
The last expression is equivalent to Kubo's formula and represents our last result. By definition, $t$ is an arbitrary time after which the collision has ended $t \geq \tau_{\mathrm{end}}$, where $\tau_{\mathrm{end}} \equiv t - |(x_{\mathrm{max}} - x_0)|/v_0$ is determined by $\lambda^{l}(\tau) = 0$. The interpretation of Eq.~\eqref{kubo} within scattering theory is clear: an incoming massive particle approaches the scattering region from the distant past, inducing an effective time-dependent interaction on the system as it travels through the collision region. This causes a response of observable $\hat{A}$ in time which lasts until the particle has gone through this region. {A numerical comparison between Eq.~\eqref{kubo} and the exact scattering dynamics is illustrated in Fig.~\ref{fig}}.

\section{Conclusions \label{sec:conclusions}}

We have shown how linear response theory for closed quantum systems can be seen as a specific limit of quantum scattering theory. In particular, we studied the {first-order contribution of the scattering map} -- usually ignored in the literature of open quantum systems -- and shown that it encodes a non-perturbative response function in terms of scattering amplitudes. This response function obeys a non-perturbative version of the fluctuation-dissipation theorem and yields the traditional linear response function for closed systems in the Born approximation of scattering theory. An essential ingredient in recovering Kubo's formula is that the incoming particle is sufficiently massive and localized in space so that its trajectory is classical. This is consistent with other findings showing that such particles play the role of a work source \cite{Jacob2022,Piccione2024}. The dynamical map in Eq.~\eqref{cptp} has a rich structure that we did not fully explore in this study, but could be used as starting point of non-perturbative and fully-quantum investigations of thermodynamics and quantum information. On the other hand, one could also study the properties of the dynamical map for the particle instead of the system or scatterer, potentially establishing connections to quantum metrology. Finally, we hope our work opens new research avenues in condensed matter physics using the ubiquitous and experimentally-relevant framework of quantum collisions. 



\section*{Acknowledgements} S.L.J. acknowledges the financial support from a Marie Skłodowska-Curie Fellowship (Grant No. 101103884). This work was supported by the EPSRC-SFI joint project QuamNESS and by SFI under the Frontier For the Future Program. J.G. is supported by a SFI Royal Society University Research Fellowship.

\appendix

\section{Linear response theory \label{app:linearresponse}}

We review linear response theory for closed quantum systems. For the sake of simplicity, this section is self-contained and we repeat some of the notation already used in main text, hoping no confusion arises. 

We consider a closed quantum system described by a time-dependent Hamiltonian $\hat{H}(t) = \hat{H}_0 + \hat{V}(t)$ where the interaction $\hat{V}(t)$ describes an external driving that changes the bare energy of the system given by $\hat{H}_0$. If $\hat{\rho}(t_0)$ is the density operator representing the quantum state of the system at time $t_0$, then the evolved state of the system at time $t$ is given by $\hat{\rho}(t) = \hat{U}(t,t_0) \hat{\rho}(t_0) \hat{U}(t,t_0)^{\dagger}$ where $\hat{U}(t,t_0) \equiv \hat{T}_{\rightarrow} \exp[(1/i\hbar)\int^{t}_{t_0} \hat{H}(t') dt']$ is the unitary evolution operator and $\hat{T}_{\rightarrow}$ the time-ordering operator. Accordingly, the value of a system observable $\hat{A}$ at time $t$ is given by $A(t) \equiv \mathrm{Tr}[\hat{A} \hat{\rho}(t)]$. In general $\hat{A}$ itself can be time-dependent, but this does not change our treatment. It is useful to adopt the interaction picture defined by $A(t) = \mathrm{Tr}[\hat{A}_I(t,t_0) \hat{\rho}_I(t,t_0)]$, where $\hat{A}_I(t,t_0) \equiv \hat{U}_0(t,t_0)\hat{A}\hat{U}_0(t,t_0)^{\dagger}$ is the observable in this picture defined by the free evolution operator $\hat{U}_0(t,t_0) \equiv e^{-i \hat{H}_0 t /\hbar}$ and $\hat{\rho}_I(t,t_0) \equiv \hat{U}_I(t,t_0)\hat{\rho}(t_0)\hat{U}_I(t,t_0)^{\dagger}$ where $\hat{U}_I(t,t_0) \equiv \hat{U}_0(t,t_0)^{\dagger} \hat{U}(t,t_0)$ is the interaction picture evolution operator. In this picture, the evolution is governed by the equation
\begin{align}
    \frac{d \hat{\rho}_I(t,t_0)}{dt} = -\frac{i}{\hbar} [\hat{V}_I(t,t_0),\hat{\rho}_I(t,t_0)] \; .
\end{align}
Integrating the last expression between $t_0$ and some larger time leads to $\hat{\rho}_I(t,t_0) = \hat{\rho}(t_0) - i \hbar^{-1} \int^{t}_{t_0} dt' [\hat{V}_I(t',t_0),\hat{\rho}_I(t',t_0)]$, which can be used as starting point of perturbation theory. Assuming a weak interaction $\hat{V}(t)$, we can insert the left hand-side on the integrand in the right hand-side and keep only the first term
\begin{align}
    \hat{\rho}_I(t,t_0) = \hat{\rho}(t_0) - \frac{i}{\hbar} \int^{t}_{t_0} dt' [\hat{V}_I(t',t_0),\hat{\rho}(t_0)] \; .
\end{align}
Substituting the last expression into $A(t) = \mathrm{Tr}[\hat{A}_I(t,t_0) \hat{\rho}_I(t,t_0)]$, assuming that the initial state is stationary with respect to free evolution $[\hat{H}_0,\hat{\rho}(t_0)]=0$ and using the cyclic property of the trace leads to
\begin{align}
    \label{linear-response}
    \delta A(t) = - \frac{i}{\hbar} \int^{t}_{t_0} dt' \mathrm{Tr}[[\hat{A}_I(t-t'),\hat{V}(t')]\hat{\rho}(t_0)] \; ,
\end{align}
where $\delta A(t) \equiv A(t) - A(t_0)$. Eq.~\eqref{linear-response} defines a linear (first-order) perturbation for an initially stationary system. However, Kubo makes two more assumptions, namely that the perturbation is of the form $\hat{V}(t) = \lambda(t) \hat{V}$ where $\lambda(t)$ is a time-dependent force function and that the perturbation starts in the infinite past $t_0 \rightarrow -\infty$ \cite{Kubo1957-I,Kubo1966}. Then Eq.~\eqref{linear-response} can be written
\begin{align}
    \label{kubo-formula}
    \delta A(t) = \int^{t}_{-\infty} dt' \phi_{AV}(t-t') \lambda(t') \; ,
\end{align}
where the response function encodes the response of observable $\hat{A}$ at time $t$ to a perturbation $\hat{V}$ at time $t=0$ 
\begin{align}
    \label{kubo-response}
    \phi_{AV}(t) \equiv -\frac{i}{\hbar} \mathrm{Tr}[[\hat{A}_I(t),\hat{V}]\hat{\rho}] \; ,
\end{align}
and $\hat{\rho} \equiv \lim_{t_0 \rightarrow -\infty} \hat{\rho}(t_0)$ denote the initial state in the infinity past (before the perturbation is applied). It is easy to show that the response function in Eq.~\eqref{kubo-formula} is homogeneous in time due to the fact that $\hat{\rho}$ is stationary. Eq.~\eqref{kubo-formula} is best studied by changing to the frequency domain through
\begin{align}
    \lambda(t) = \frac{1}{2 \pi} \int^{+\infty}_{-\infty} d\omega~\tilde{\lambda}(\omega)e^{-i \omega t} \; ,
\end{align}
where $\tilde{\lambda}(\omega)$ is the Fourier transform of $\lambda(t)$. Then Eq.~\eqref{kubo-formula} is written as
\begin{align}
    \delta A(t) = \frac{1}{2 \pi} \int^{+\infty}_{-\infty} d \omega~\chi_{AV}(\omega) \tilde{\lambda}(\omega)e^{-i \omega t} \; ,
\end{align}
where the susceptibility is defined by
\begin{align}
    \label{susceptibility}
    \chi_{AV}(\omega) \equiv \int^{+ \infty}_{0} \phi_{AV}(t) e^{i \omega t} \; .
\end{align}
Note that the susceptibility is the Fourier transform of the response function $\phi^{\rightarrow}_{AV}(t) \equiv \theta(t) \phi_{AV}(t)$, where $\theta(t)$ is the Heaviside Theta function enforcing causality. The susceptibility is one of the most important quantities in linear response theory and a fluctuation-dissipation theorem can be derived assuming $\hat{\rho} = \hat{\omega}_{\beta}$ where is the thermal state $\hat{\omega}_{\beta} \equiv e^{-\beta \hat{H}_0} / Z$ and $Z = \mathrm{Tr}[e^{-\beta \hat{H}_0}]$ is the partition function and $\beta \equiv 1/k_B T$ where $k_B$ is Boltzmann's constant and $T$ is the temperature. Introducing the symmetric correlation function
\begin{align}
    C_{AV}(t) \equiv \frac{1}{2}\mathrm{Tr}[\{\hat{A}_I(t),\hat{V} \}\hat{\omega}_{\beta}] \; ,
\end{align}
we can see that its Fourier transform can be written
\begin{align}
    C_{AV}(\omega) & \equiv \int^{+\infty}_{-\infty} C_{AV}(t) e^{i \omega t} \nonumber \\
    \label{correlation}
    & = (1 + e^{-\beta \hbar \omega}) \pi \hbar \sum_{m,n} \braket{m|\hat{A}|n} \braket{n|\hat{V}|m} \frac{e^{-\beta E_m}}{Z} \delta(E_m - E_n + \hbar \omega) \; ,
\end{align}
where we used the eigenbasis of the bare Hamiltonian $\{ \ket{m} \}$ defined by $\hat{H}_0 \ket{m} = E_m \ket{m}$ and the integral representation of the $\delta$ function already presented in Eq.~\eqref{deltafunction}. A similar exercise can be done for Eq.~\eqref{susceptibility} which leads to
\begin{align}
    \chi_{AV}(\omega) & = \sum_{n,m} \braket{m|\hat{A}|n} \braket{n|\hat{V}|m} \frac{e^{-\beta E_m} - e^{-\beta E_n}}{Z}  \bigg[ \mathcal{P} \frac{1}{E_m - E_n + \hbar \omega} - i \pi \delta(E_m - E_n +  \hbar \omega) \bigg] \nonumber \\
    & = -i \pi (1 - e^{-\beta \hbar \omega})\sum_{n,m} \braket{m|\hat{A}|n} \braket{n|\hat{V}|m}\frac{e^{-\beta E_m}}{Z} \delta(E_m - E_n +  \hbar \omega) + \sum_{n,m} \braket{m|\hat{A}|n} \braket{n|\hat{V}|m} \frac{e^{-\beta E_m} - e^{-\beta E_n}}{Z}  \mathcal{P} \frac{1}{E_m - E_n + \hbar \omega} \nonumber \\
    \label{susceptibility-decomposition}
    & = -\frac{i}{\hbar} \tanh\bigg(\frac{\beta \hbar \omega}{2}\bigg) C_{AV}(\omega) + \sum_{n,m} \braket{m|\hat{A}|n} \braket{n|\hat{V}|m} \frac{e^{-\beta E_m} - e^{-\beta E_n}}{Z}  \mathcal{P} \frac{1}{E_m - E_n + \hbar \omega} \; .
\end{align}
In the first line we used the equality $\int_{0}^{+\infty} dx~e^{ \pm i x a} = \pm i [\mathcal{P} a^{-1} \mp i \pi \delta(a)]$ where $\mathcal{P}$ denotes the Cauchy principal value. In the second and third lines, we split the contributions and identified the Fourier transform of the correlation function in Eq.~\eqref{correlation}. For $\hat{A} = \hat{V}$ the imaginary part of the response can be written
\begin{align}
    \label{fdr}
    \Im[\chi_{AA}(\omega)] = -\frac{1}{\hbar} \tanh\bigg(\frac{\beta \hbar \omega}{2}\bigg) C_{AA}(\omega) \; ,
\end{align}
which is the well-known form of the fluctuation-dissipation relation connecting the response to the auto-correlation function. {More generally, an equivalent of Eq.~\eqref{fdr} also holds for $\hat{A} \neq \hat{V}$ when the second term in Eq.~\eqref{susceptibility-decomposition} is purely real.}

\section{Scattering properties \label{app:scattmatrixproperties}} 

In this section, we present several properties which follow from the unitarity of the scattering operator $\hat{S}^{\dagger}\hat{S} = \hat{S} \hat{S}^{\dagger} = \hat{\mathbb{I}}$, where $\hat{\mathbb{I}} = \hat{\mathbb{I}}_S \otimes \hat{\mathbb{I}}_P$. Unitarity enforces two properties on the scattering matrix which can be obtained through the expressions 
\begin{align}
    \braket{E^{\alpha'}_{p'},j'|\hat{S}^{\dagger}\hat{S}|E^{\alpha}_{p},j} & = \delta_{\alpha \alpha'} \delta_{j'j} \delta(E_p-E_{p'}) \\
    \braket{E^{\alpha'}_{p'},j'|\hat{S} \hat{S}^{\dagger}|E^{\alpha}_{p},j} &= \delta_{\alpha \alpha'} \delta_{j'j} \delta(E_p-E_{p'}) \; .
\end{align}
By inserting resolutions of identity in the full Hilbert space between the two operators, using the representation of the scattering operator in Eq.~\eqref{scattopmatrix} and integrating out one of the $\delta$ functions, we find that the scattering matrix obeys
\begin{align}
    \sum_{\beta,k} [s^{\beta \alpha'}_{kj'}(E)]^* s^{\beta \alpha}_{kj}(E) & = \delta_{\alpha \alpha'} \delta_{j'j} \label{scattprop1}\\
     \sum_{\beta,k} s^{\alpha' \beta}_{j'k}(E) [s^{\alpha \beta}_{jk}(E)]^* & = \delta_{\alpha \alpha'} \delta_{j'j} \label{scattprop2}\; ,
\end{align}
where $E = E_p + e_j$ is the total energy. These are the most important properties of the scattering matrix holding for any fixed total energy $E$. They imply that the transition probability $P^{\alpha' \alpha}_{j'j}(E) \equiv |s^{\alpha'\alpha}_{j'j}(E)|^2$ obeys $\sum_{j',\alpha'} P^{\alpha' \alpha}_{j'j}(E) = \sum_{j,\alpha} P^{\alpha' \alpha}_{j'j}(E) = 1$. 

As we have shown in the main text, the scattering operator can be decomposed as $\hat{S} = \hat{\mathbb{I}} - i \hat{T}$ and the unitary property then yields $i(\hat{T} - \hat{T}^{\dagger}) = \hat{T}^{\dagger} \hat{T}$ where $\hat{T}$ is a normal operator. This property, at the level of the corresponding scattering amplitude defined by Eq.~\eqref{scattmatrix-decomposition}, can be obtained from Eq.~\eqref{scattprop1}
\begin{align}
    \label{optical-general}
    i ( t^{\alpha' \alpha}_{j'j}(E) - [t^{\alpha \alpha'}_{jj'}(E)]^* ) = \sum_{\beta,k} [t^{\beta \alpha'}_{kj'}(E)]^* t^{\beta \alpha}_{kj}(E) \; .
\end{align}
A similar expression can be obtained from Eq.~\eqref{scattprop2}. Eq.~\eqref{optical-general} is a general version of the optical theorem of scattering theory \cite{Belkic2004,Taylor2006}. Choosing $j=j'$ and $\alpha = \alpha'$ it becomes
\begin{align}
    \label{optical-specific}
    \Im[t^{\alpha \alpha}_{jj}(E)] = -\frac{1}{2} \sum_{\beta,k} |t^{\beta \alpha}_{kj}(E)|^2 \equiv - \frac{1}{2}\sigma^{\alpha}_{j}(E) \leq 0 \; .
\end{align}
We thus obtain a relationship between the imaginary part of the elastic component of the scattering amplitude and the cross-section $\sigma^{\alpha}_j(E)$ at total energy $E$ for a system initially in state $\ket{j}$ and particle travelling with direction $\alpha$. The optical theorem also shows that the imaginary part is necessarily negative. Note that the cross-section is an adimensional quantity in one dimension, while it has the dimension of length and area in two and three dimensions, respectively. 

Finally, just like we defined the eigenoperators in Eq.~\eqref{eigenoperators}, the decomposition in Eq.~\eqref{scattmatrix-decomposition} allow us to define 
\begin{align}
    \hat{S}^{\alpha' \alpha}_{\Delta}(E_p) \equiv \sum_{\substack{j',j:\\ e_{j'}-e_j = \Delta}} s_{j'j}^{\alpha' \alpha}(E_p + e_j) \ket{j'}\bra{j} = \delta_{\Delta,0}\mathbb{I}_S - i \hat{T}^{\alpha' \alpha}_{\Delta}(E_p) \; .
\end{align}
Using Eq.~\eqref{scattprop1}, it is easy to show that 
\begin{align}
    \sum_{\Delta, \alpha''}\hat{S}^{\alpha'' \alpha'}_{\Delta}(E_p)^{\dagger} \hat{S}^{\alpha'' \alpha}_{\Delta}(E_p) = \delta_{\alpha \alpha'}\mathbb{I}_S
\end{align}
which immediately leads to
\begin{align}
    i[\hat{T}^{\alpha' \alpha}_0(E_p) - \hat{T}^{\alpha \alpha'}_0(E_p)^{\dagger}] = \sum_{\Delta,\alpha''} \hat{T}^{\alpha'' \alpha'}_{\Delta}(E_p)^{\dagger} \hat{T}^{\alpha'' \alpha}_{\Delta}(E_p) \; . 
\end{align}
This is an expression of the optical theorem in Eq.~\eqref{optical-general} at the level of eigenoperators. Note that when $\alpha = \alpha'$ the operator on the right hand-side becomes positive $\hat{\sigma}^{\alpha}(E_p) \equiv \sum_{\Delta,\alpha''} \hat{T}^{\alpha'' \alpha}_{\Delta}(E_p)^{\dagger} \hat{T}^{\alpha'' \alpha}_{\Delta}(E_p) \geq 0$ and can be interpreted in terms of a cross-section. This operator plays a role in the dissipative part of the system evolution as can be seen from Eq.~\eqref{dissipator} or Eq.~\eqref{dissipator-narrow}.

{\section{Quantum master equation \label{app:qme}}

In this section, we show how to obtain a quantum master equation from the scattering map in Eq.~\eqref{cptp}, defined by Eq.~\eqref{lambshift} and \eqref{dissipator}. In doing so, we establish a clearer connection between the scattering map -- which is exact and describes dynamics in discrete time -- and the more familiar quantum master equations.

Consider that, after the collision with the particle, the quantum system $S$ evolves freely for some time $t$, at which point a second collision happens. Introducing the superoperator of free evolution $\mathcal{U}_t ( \cdot ) \equiv e^{-i \hat{H}_S t /\hbar} \cdot e^{i \hat{H}_S t /\hbar}$, we can define the state of the system at time $t$ after the first collision as
\begin{align}
    \hat{\rho}^{1}_S(t) \equiv (\mathcal{U}_t\circ \Phi)(\hat{\rho}_S) \; .
\end{align}
The crucial assumption is to consider that the waiting times are random. If $g(t)$ denotes the probability distribution for these waiting times, then the average state of the system at time $t$ after the first collision is given by
\begin{align}
     \hat{\bar\rho}^{1}_S(t) = \int^{t}_{0} d\tau~ g(t-\tau) \hat{\rho}^{1}_S(\tau) \; .
\end{align}
After the second collision, then the system evolves freely again for some random time, and the process goes on for $n$ collisions. The average state of the system after $n$ collisions is then
\begin{align}
     \hat{\bar\rho}^{n}_S(t) = \int^{t}_{0} d\tau~ g(t-\tau) (\mathcal{U}_{t-\tau}\circ \Phi)(\hat{\bar\rho}^{n-1}_S(\tau)) \; .
\end{align}
Assuming that the waiting times are Poissonian $g(t) = \gamma e^{-\gamma t}$ -- where $\gamma$ is the rate of collisions, or the inverse of the average waiting time -- and differentiating the last expression with respect to time, we have
\begin{align}
    \frac{d  \hat{\bar\rho}^{n}_S(t)}{dt} = -\frac{i}{\hbar} [\hat{H}_S, \hat{\bar\rho}^{n}_S(t)] + \gamma[ \Phi(\hat{\bar\rho}^{n-1}_S(t)) - \hat{\bar\rho}^{n}_S(t)] \; .
\end{align}
The last step involves an averaging over all possible number of collisions, so we obtain a state that is unconditioned in $n$, i.e. $\langle \hat{\bar\rho}^{n}_S(t) \rangle \equiv \hat{\bar\rho}_S(t)$. Using the explicit form of the map in Eq.~\eqref{cptp}, we obtain Eq.~\eqref{qme}. More details on the derivation and use of this quantum master equation can be found in Refs.~\cite{Karplus1948,Strasberg2017,Tabanera2023}.

}

\section{Fourier transform of the response \label{app:fouriertransform}} 

In this section, we express our results using the more well-known continuous Fourier transform. Let us express Eq.~\eqref{chiA-2} in the frequency domain by introducing the Fourier transform of the particle state
\begin{align}
    \rho_P^{\alpha \alpha'}(E_p,E_{p'},t) = \frac{1}{2 \pi} \int^{+\infty}_{-\infty} d\omega~\rho_P^{\alpha \alpha'}(E_p,E_{p'},\omega) e^{-i \omega t} \nonumber \; ,
\end{align}
so that we obtain
\begin{align}
    \chi_{A} = \frac{1}{2 \pi} \int^{+\infty}_{-\infty} d\omega \int dE_p \int dE_{p'} 
 \sum_{\alpha',\alpha}~\chi_A^{\alpha' \alpha}(E_p,\omega) \rho_P^{\alpha \alpha'}(E_p,E_{p'},\omega) \; .
\end{align}
In the last expression we have identified the Fourier transform of the response function in Eq.~\eqref{response-time}
\begin{align}
    \label{response-omega}
    \chi_A^{\alpha' \alpha}(E_p,\omega) = \int^{+\infty}_{-\infty} dt~e^{i \omega t} \chi_A^{\alpha' \alpha}(E_p,t) = \sum_{\Delta} \delta(\hbar \omega - \Delta) \chi_{\Delta}^{\alpha' \alpha}(E_p) \; .
\end{align}
The last expression clarifies the relationship between the discrete Fourier transform used in the main text and the continuous Fourier transform. In the same way, we can use Eq.~\eqref{correlation-time} to show that
\begin{align}
    \label{correlation-omega}
    C_A^{\alpha' \alpha}(E_p,\omega) = \int^{+\infty}_{-\infty} dt~e^{i \omega t} C_A^{\alpha' \alpha}(E_p,t) = \hbar\sum_{\Delta} \delta(\hbar \omega - \Delta) C_{\Delta}^{\alpha' \alpha}(E_p)
\end{align}
Henceforth assuming a thermal state and using Eqs.~\eqref{fdr-Delta} and \eqref{correlation-omega} we immediately obtain
\begin{align}
    \chi_A^{\alpha' \alpha}(E_p,\omega) 
    \label{fdr-omega}
    & = - \frac{2i}{\hbar}  \tanh\bigg(\frac{\beta \hbar \omega}{2}\bigg) C_A^{\alpha'\alpha}(E_p,\omega) \; ,
\end{align}
which shows that the fluctuation-dissipation relation in Eq.~\eqref{fdr-Delta}, expressed in terms of a discrete Fourier transform, also holds for its continuous counterpart. We note that in opposition to the susceptibility in Eq.~\eqref{susceptibility} for closed systems, the integral in Eq.~\eqref{response-omega} runs over all time. However, let us split Eq.~\eqref{response-omega} into two components $\chi_A^{\alpha' \alpha}(E_p,\omega) = {}^{+} \chi_A^{\alpha' \alpha}(E_p,\omega) + {}^{-} \chi_A^{\alpha' \alpha}(E_p,\omega)$ where
\begin{align}
    {}^{\pm} \chi_A^{\alpha' \alpha}(E_p,\omega) = \int^{+\infty}_{0} dt~e^{ \pm i \omega t} \chi_A^{\alpha' \alpha}(E_p, \pm t) \; .
\end{align}
The positive component is a non-perturbative version of the susceptibility in Eq.~\eqref{susceptibility} and captures the causal (retarded) component of the response, while the negative component captures the non-causal (advanced) component. Writing the response in the integrand in terms of the discrete Fourier transform in Eq.~\eqref{response-time} we have
\begin{align}
    {}^{\pm} \chi_A^{\alpha' \alpha}(E_p,\omega) & = \frac{1}{2 \pi \hbar} \sum_{\Delta} \chi_{\Delta}^{\alpha' \alpha}(E_p) \int^{+\infty}_{0} dt~e^{ \pm i (\hbar \omega - \Delta) t/\hbar} \nonumber \\
    & = \frac{\pm i}{2 \pi} \sum_{\Delta} \chi_{\Delta}^{\alpha' \alpha}(E_p) \bigg[ \mathcal{P} \frac{1}{\hbar \omega - \Delta} \mp i \pi \delta(\hbar \omega - \Delta) \bigg] \nonumber \\
    & = \frac{1}{2} \sum_{\Delta} \chi_{\Delta}^{\alpha' \alpha}(E_p) \delta(\hbar \omega - \Delta) \pm  \frac{i}{2 \pi} \sum_{\Delta} \chi_{\Delta}^{\alpha' \alpha}(E_p) \mathcal{P} \frac{1}{\hbar \omega - \Delta} \nonumber \\
    & = -\frac{i}{\hbar} \tanh\bigg(\frac{\beta \hbar \omega}{2}\bigg) C_A^{\alpha' \alpha}(E_p,\omega) \pm \frac{i}{2 \pi} \sum_{\Delta} \chi_{\Delta}^{\alpha' \alpha}(E_p) \mathcal{P} \frac{1}{\hbar \omega - \Delta} \; . 
\end{align}
In the second line, we used Cauchy's principal value theorem already used in Eq.~\eqref{susceptibility-decomposition}. In the third line, we split into two contributions. In the last line, we used again Eq.~\eqref{fdr-Delta} and Eq.~\eqref{correlation-omega}. Note that the first term in the last expression is the non-perturbative equivalent of the first term in Eq.~\eqref{susceptibility-decomposition}. On the other hand, in this non-perturbative regime, we cannot simplify the last expression to yield auto-correlations functions obeying Eq.~\eqref{fdr}.

\bibliography{references}

\end{document}